\documentclass[sn-basic]{sn-jnl}

\jyear{2023}%

\theoremstyle{thmstyleone}%
\theoremstyle{thmstyletwo}%

\theoremstyle{thmstylethree}%

\chardef\us=`\_

\usepackage{amsmath}

\begin{document}

\title[The solar activity-solar wind predictive causality at Space Climate scales]{Disentangling the solar activity-solar wind predictive causality at Space Climate scales}


\author*[1]{\fnm{Raffaele} \sur{Reda}}\email{raffaele.reda@roma2.infn.it}

\author[2]{\fnm{Mirko} \sur{Stumpo}}\email{mirko.stumpo@inaf.it}

\author[1]{\fnm{Luca} \sur{Giovannelli}}\email{luca.giovannelli@roma2.infn.it}

\author[3,2]{\fnm{Tommaso} \sur{Alberti}}\email{tommaso.alberti@ingv.it}

\author[2]{\fnm{Giuseppe}
\sur{Consolini}}\email{giuseppe.consolini@inaf.it}

\affil[1]{\orgdiv{Department of Physics}, \orgname{University of Rome Tor Vergata}, \orgaddress{\street{Via della Ricerca Scientifica 1}, \city{Rome}, \postcode{00133}, \country{Italy}}}

\affil[2]{\orgname{INAF - Istituto di Astrofisica e Planetologia Spaziali}, \orgaddress{\street{Via del Fosso del Cavaliere 100}, \city{Rome}, \postcode{00133}, \country{Italy}}}

\affil[3]{\orgname{Istituto Nazionale di Geofisica e Vulcanologia}, \orgaddress{\street{Via di Vigna Murata 605}, \city{Rome}, \postcode{00143}, \country{Italy}}}


\abstract{The variability in the magnetic activity of the Sun is the main source of the observed changes in the plasma and electromagnetic environments within the heliosphere. 
The primary way in which solar activity affects the Earth's environment is via the solar wind and its transients.
However, the relationship between solar activity and solar wind is not the same at the Space Weather and Space Climate time scales.
In this work, we investigate this relationship exploiting five solar cycles data of Ca II K index and solar wind parameters, by taking advantage of the Hilbert-Huang Transform, which allows to separate the contribution at the different time scales. By filtering out the high frequency components and looking at decennial time scales, we confirm the presence of a delayed response of solar wind to Ca II K index variations, with a time lag of $\sim$ 3.1-year for the speed and $\sim$ 3.4-year for the dynamic pressure. To assess the results in a stronger framework, we make use of a Transfer Entropy approach to investigate the information flow between the quantities and to test the causality of the relation. The time lag results from the latter are consistent with the cross-correlation ones, pointing out the presence of a statistical significant information flow from Ca II K index to solar wind dynamic pressure that peaks at time lag of 3.6-year. Such a result could be of relevance to build up a predictive model in a Space Climate context.}




\maketitle

\section{Introduction}
\label{introduction_section}

The presence of a magnetic field in the solar atmosphere is the most prominent manifestation of solar variability. Observations of such magnetic variability on the Sun can provide strong observational constraints on the solar dynamo theory, helping to understand the physical mechanisms underlying magnetic flux emergence and evolution. This is particularly interesting on long-term scales, as the solar cycle can offer insights into the complex dynamics of the global dynamo \citep[e.g.][]{Usoskin2007}.

The goal of Space Climate is to describe long-term variations in solar activity and their impact on the heliosphere and the Earth's environment. The time scale usually used to distinguish the two regimes (i.e., Space Weather and Space Climate) is a few solar rotations \citep{Mursula2007}.
Here, we focus on the multi-year time scale relationship between solar activity and near-Earth solar wind properties. To achieve this goal, we utilize the full extent of space-age solar wind observations via the OMNI database \citep{King2005}. Furthermore, we make use of a century-long dataset of the Ca II K index, a widely employed physical solar activity indicator. This emission index has been demonstrated to be a reliable proxy for the magnetic flux density at the Sun \citep{Schrijver1989, Ortiz2005, Chatzistergos2019a} and it has also been proven to trace long-term variations in solar activity \citep[e.g.][]{Judge2006,Bertello2016,Chatzistergos2019b}.

The first evidence of a strong link between solar wind properties and the solar cycle comes from observations of a period close to 11 years since the very first satellite observations \citep{Siscoe1978,King1979,Neugebauer1981}.
In later studies, long-term solar wind properties were mostly compared with sunspot numbers, revealing a non-perfect match among the periodicities of the solar cycle and solar wind speed, density, pressure, and magnetic field, further highlighting a delayed response of solar wind signals compared to sunspot numbers \citep{Petrinec1991, Kohnlein96, ElBorie02, Katsavrias12, Richardson12, Li2016, Li2017, Venzmer2017, Samsonov19}.
A recent observational analysis of near-Earth solar wind measurements in relation to the Ca II K index was the first to study these properties over the last five solar cycles \citep{Reda2023}. A 3.2-year lag of solar wind speed with respect to the Ca II K index is found using both cross-correlation and mutual information analysis, while a 3.6-year lag is found between the magnetic proxy and solar wind dynamic pressure.
The analysis on the time lag behaviour between the Ca II K index and the same solar wind parameters was further extended in \cite{Reda2023AdSpR}, studying how their pairwise relative lags vary over the last five solar cycles, with values ranging from 6 years to 1 year.

In \cite{Reda2023AdSpR} and \cite{Reda2023}, the Space Climate scales were studied by applying a 37-month moving average to the Ca II K index and solar wind parameters, following the approach used by \cite{Kohnlein96}.
However, the use of the moving average as a low-pass filter can remove some relevant features at the Space Climate scales, such as the double maxima present in some solar cycles. In this study, we propose to overcome these issues by using the Hilbert-Huang Transform \citep{Huang98}, and in particular the Empirical Mode Decomposition, to filter out the intrinsic modes with mean periods below 3 years. We, therefore, reproduce a similar analysis as in other studies, studying the delays between the signals using a cross-correlation analysis.

Furthermore, we aim to investigate the causal link between the proxy of solar activity and the characteristics of the solar wind. This causal connection can guide us in exploring the underlying physical mechanisms responsible for this relationship, opening significant prospects for understanding the mechanisms of the solar dynamo at Space Climate scales.
For this reason, we intend to use the Transfer Entropy \citep{Schreiber2000}, a novel approach from information theory recently applied to the analogous complex dynamics of the Earth's magnetosphere-ionosphere system \citep{Stumpo2020, balasis2023complex, stumpo2023dynamical}. Transfer Entropy can track down the information flow between variables in different directions, thus showing the causality relation between the variables.
Although a causal relationship between the driver of solar variability and solar wind properties is expected, such a measure is not assured and can shed light on the parameters of the solar wind that are more influenced. Additionally, this type of analysis provides an independent measure of the solar wind's response times to solar magnetic variability, which is not constrained by a linear analysis. In particular, in this study, we aim to reanalyse the unfiltered data, sampled monthly, to study the solar wind properties on climatological time scales.

These investigations are crucial in an era of significant expansion of human activity in space. The variations of the space radiation environment, for example, have been forecasted for the next 80 years in the perspective of long-term changes in the space climate \citep{Barnard2011}. We are at the beginning of a new era of human exploration of deep space, which also aims to colonize and inhabit environments unprotected by the Earth's magnetosphere. Solar wind constitutes the main low-energy and high-flux component of charged particles in the space environment and has a significant impact on the possibility of permanently inhabiting remote locations in the solar system, such as the lunar surface in the near future and Mars in the medium term. For this reason, it is essential to understand the physical mechanisms governing the variability of the solar wind on multi-year scales.

The present article is structured as follows: Section \ref{data section} gives a description of the data used; Section \ref{method section} explains the techniques adopted for the analysis here performed; Section \ref{results section} presents the results of the analysis; finally, in Section \ref{discussion and conclusions section} we list and discuss the results obtained.

\section{Data}
\label{data section}
The data we use in the present study, to investigate the relationship between the solar magnetic activity and the near-Earth solar wind, are measurements of the Ca II K index and of the solar wind speed and dynamic pressure.
In particular, we use here the monthly averages of the parameters listed above over the period 1965-2021.

The Ca II K index is a physical proxy of solar magnetic activity that accounts for the emission in the K line of Ca II at 393.4 nm. Such a line is originated in the middle solar atmosphere (i.e., the chromosphere) and it is related to the mean chromospheric emission of the Sun. The Ca II K index is one of the more commonly used activity indices and it has been proven to be a great proxy for the line-of-sight (LoS) unsigned magnetic flux density along all the phases of the cycle, and not only when sunspots are present \cite[see e.g.][]{Schrijver1989, Ortiz2005, Chatzistergos2019a}. Specifically, in this work we make use of the Ca II K index composite presented and described in \cite{Bertello2016}, which is freely accessible from the National Solar Observatory (NSO) website at \url{https://solis.nso.edu/0/iss/}. The Ca II K 0.1 nm emission index contains inter-calibrated measures from three different observatories (Kodaikanal Solar Observatory, Sacramento Peak and ISS-SOLIS at NSO) and overall it covers the time period between February 1907 and October 2017. After that date, the SOLIS facility has been offline and no more data from this instrument are available. However, this dataset has been already extended to April 2021 in \cite{Reda2023}, by making use of the Mg II index (University of Bremen composite), which has been proven to strongly correlate with the Ca II K index \cite[see e.g.][]{Donnelly1994,Reda2021,Reda2023}.

The near-Earth solar wind data are taken from the OMNI database, which can be accessed at \url{https://omniweb.gsfc.nasa.gov/hw.html}. The OMNI dataset is a collection of various near-Earth solar wind parameters, both magnetic and plasma ones, provided with different time resolutions \citep{King2005}. It is compiled by using validated data from several spacecrafts, such as IMP, ISEE, ACE, Wind and Geotail. Among the set of parameters provided by the OMNI database, the analysis we perform in this study regards two dynamic parameters of the solar wind: speed ($\mathrm{V_{sw}}$) and dynamic pressure ($\mathrm{P_{d,sw}}$). The latter has been computed starting from the speed ($\mathrm{V_{sw}}$) and the ion density ($\mathrm{n_{i,sw}}$), as $\mathrm{P_{d,sw} = 1/2\,m_{p} n_{i,sw} V_{sw}^2}$, where the proton mass $\mathrm{m_{p}}$ is assumed as the mean ion mass. Data concerning these parameters
are available starting from July 1965, thus constituting the main limit for the temporal extension of the analysis we carry out here.

The missing monthly data of Ca II K index, solar wind speed and dynamic pressure were filled by using a simple linear interpolation between the previous and the following monthly data. This procedure allows us to continuously investigate, in the present study, the time interval that goes from July 1965 to April 2021, ensuring to almost fully cover the solar cycles from 20 to 24, together with the beginning of solar cycle 25.

\begin{figure}
    \centering
    \includegraphics[width=0.8\textwidth]{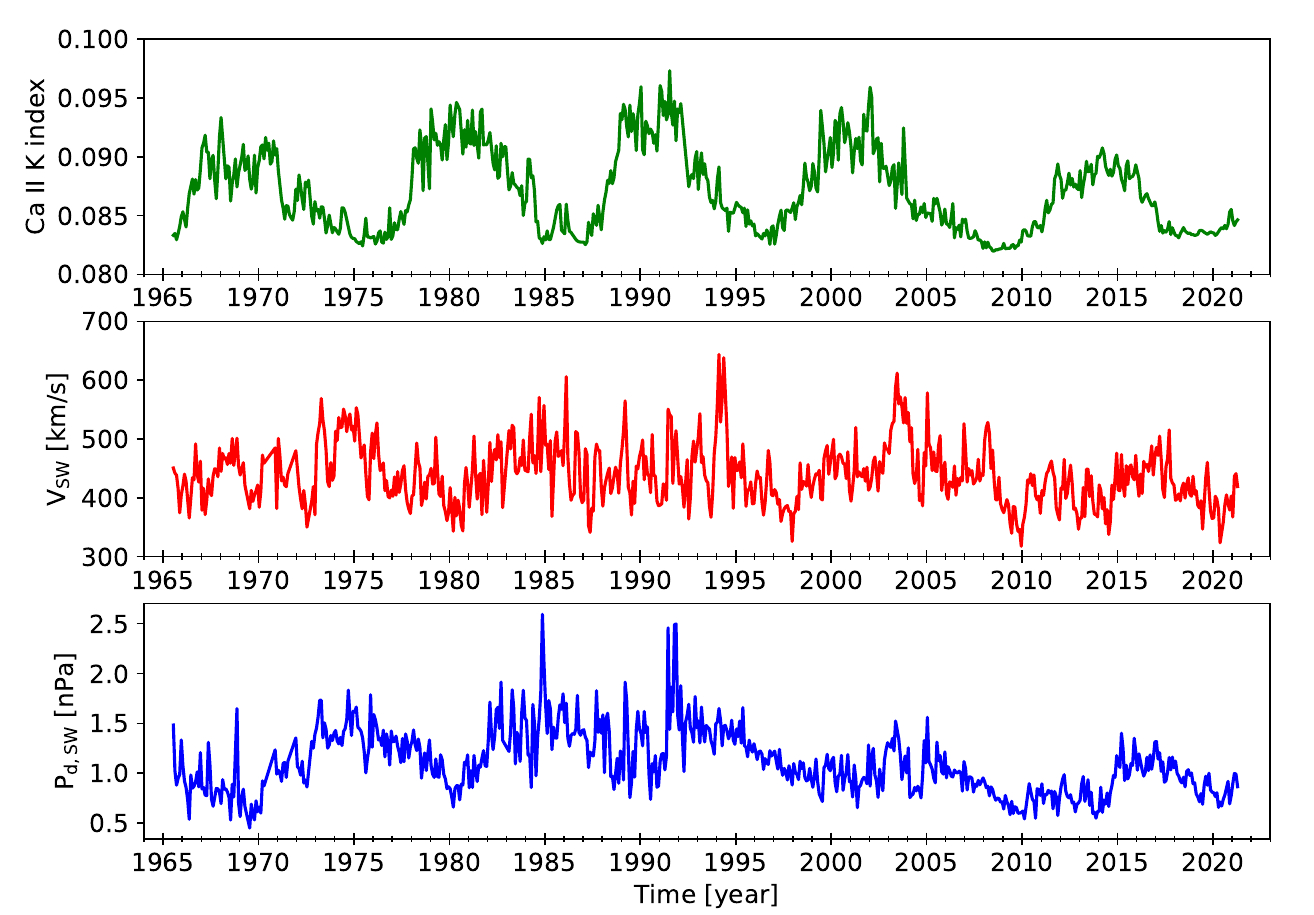}
    \caption{Monthly averages of the timeseries used for this work: Ca II K index (green, top), solar wind speed (red, middle) and solar wind dynamic pressure (blue, bottom).}
    \label{Data}
\end{figure}

\section{Methods}
\label{method section}

\subsection{The Hilbert-Huang Transform: Empirical Mode Decomposition and Hilbert Spectral Analysis}
\label{sec:HHT}

The Empirical Mode Decomposition (EMD), e.g., the first step of the Hilbert-Huang Transform (HHT), has been firstly introduced by \cite{Huang98} as an adaptive and a posteriori decomposition method whose decomposition basis is derived via an iterative process, known as sifting process, based on the local properties of signals \citep{Huang98}. 

Let $y(t)$ be a time-dependent signal, the EMD allows us to write
\begin{equation}
 y(t) = \sum_{k=1}^N c_k(t) + r(t)
 \label{eq:emd}
\end{equation}
where the set $\{c_k(t)\}$, named as Intrinsic Mode Functions (IMFs) or empirical modes, forms the decomposition basis, while $r(t)$ is the residue of the decomposition. 
The latter is a non-oscillating function of time, while an IMF is defined as a function having the same (or differing at most by one) number of extrema and zero crossings and a zero-average mean envelope derived from local maxima and minima envelopes, obtained by interpolating them by using a cubic spline \citep[e.g.,][]{Huang98,Huang08}. 
Table \ref{tab1} summarizes the main steps of the sifting process.
\begin{table}[h]
\caption{The main steps of the sifting process.}
\label{tab1}
\centering 
\begin{tabular}{l}
\hline
\hline
$y(t) \rightarrow y_m(t) = y(t) - \langle y(t)\rangle$ \\
$\delta(t) = y_m(t)$ \\
\hline
1. find local extrema of $\delta(t)$ \\ 
2. find upper and lower envelopes by using cubic spline $\rightarrow \mathcal{U}(t) \, , \, \mathcal{L}(t)$ \\ 
3. find the mean envelope $\rightarrow \mathcal{M}(t) = \frac{\mathcal{U}(t) + \mathcal{L}(t)}{2}$ \\ 
4. update $\delta(t) \rightarrow \delta(t) - \mathcal{M}(t)$ \\
\hline
{\bf if} $\delta(t)$ is an IMF \\
store $c_k(t) = \delta(t)$ \\
$\delta(t) \to \delta(t) = y_m(t) - \delta(t)$ \\
repeat steps 1.-4. \\
{\bf else} \\
iterate steps 1.-4. until $\delta(t)$ is an IMF \\
store $c_k(t) = \delta(t)$ \\
$\delta(t) \to \delta(t) = y_m(t) - \delta(t)$ \\
repeat steps 1.-4. \\
\hline
stop the process when $r(t) = \delta(t)$ is a non-oscillating function or has only two extrema \\
\hline
\hline
\end{tabular}
\end{table}

The authors proposed in \cite{Huang98} the following constraints as exit condition to stop the sifting process
\begin{equation}
    \sigma_{n} = \sum_{j} \frac{\left[\delta_{n}(t_j) - \delta_{n+1}(t_j)\right]^2}{\delta_{n}(t_j)^2} < \epsilon,
\end{equation}
being fixed $\epsilon \in [0.2,0.3]$. This criterion has been refined by \cite{Rilling03} by the so-called threshold method based on two thresholds, $\theta_1$ and $\theta_2$, to guarantee globally small fluctuations \citep[as in][]{Huang98} and to avoid locally large excursions \citep{Flandrin04}. 

In this way, a completely adaptive procedure is built, allowing us in deriving embedded oscillations without assuming linearity and/or stationarity. The derived set of empirical modes $\{c_k(t)\}$ satisfies mathematical requirements of completeness, convergence, and local orthogonality by construction \citep{Huang98}, while global orthogonality is a posteriori guaranteed since $\langle c_k, c_{k^\prime} \rangle = \delta_{kk^\prime}$, being $\langle \dots \rangle$ the scalar product between functions, and $\delta_{kk^\prime}$ the Kronecker tensor \citep[e.g.,][]{Huang08}.

Being derived the set of empirical modes, by means of the so-called Hilbert Transform (HT), i. e. the second step of the HHT, we can write each of them as modulated both in amplitude and in frequency \citep[e.g.,][]{Huang98}. Indeed, given an empirical mode $c_k(t)$ we can define its Hilbert Transform $\hat{c}_k(t)$ as
\begin{equation}
 \hat{c}_k(t) = \frac{1}{\pi} \mathcal{P} \int_0^\infty \frac{c_k(t')}{t - t'} dt'
\end{equation}
where $\mathcal{P}$ is the Cauchy principal value. By introducing the complex signal
\begin{equation}
 \zeta_k(t) = c_k(t) + i \, \hat{c}_k(t) = \alpha_k(t) e^{i \, \varphi_k(t)}
\end{equation}
%
it follows
\begin{eqnarray}
 \alpha_k(t) &=& \sqrt{c_k(t)^2 + \hat{c}_k^2} \\
 \varphi_k(t) &=& \tan^{-1} \left[ \frac{\hat{c}_k(t)}{c_k(t)} \right]
\end{eqnarray}
where $\alpha_k(t)$ and $\varphi_k(t)$ are the instantaneous amplitude and phase of the $k-$th empirical mode, respectively. The definition of instantaneous frequency derives from the instantaneous phase as $\omega_k(t) = \frac{1}{2 \pi}\frac{d \varphi_k(t)}{dt}$. Similarly, the mean time scale is $\tau_k = \langle \omega_k^{-1}(t) \rangle_t$, with $\langle \dots \rangle_t$ identifying the time average.

\subsection{Transfer Entropy}
\label{transfer entropy section}
The notion of cause-effect is a delicate question when data from controlled experiments are not available. This is the case of complex systems in general: when dealing with a system whose complete set of dynamical variables is not known \textit{a priori} and the state of the system is monitored by some indices (which work as proxies) derived empirically, correlation may be confused with causation. 

Data-driven methods for studying the degree of causation have been developed in the recent years. Generally, these methods are based on the notion of predictability, i.e., it is said that X drives Y if the knowledge of X’s past gives us information about Y’s future, but not \textit{vice-versa}. This type of causality is known as predictive causality, and it is restricted to only two variables, X and Y respectively. 
Mathematically, the concept of predictive causality is expressed through conditional independence, i.e., it is reasonable to assume that X does not drive Y if
\begin{equation}\label{eq:gen_markov_cond}
    p(Y_t \vert \mathbf{Y}_{t-1}^{(k)}; \mathbf{X}_{t-\tau}^{(l)}) = p(Y_t \vert \mathbf{Y}_{t-1}^{(k)}),
\end{equation}
where $\mathbf{X}_{t-\tau}^{(l)} = \left( X_{t-\tau}, ..., X_{t-\tau-l}\right)$, $\mathbf{Y}_{t-1}^{(k)} = \left( Y_{t-1}, ..., Y_{t-1-k}\right)$, $p(\cdot)$ denotes the probability and $\tau$ is a time lag.
Therefore, to measure predictive causality the idea is to test Equation \eqref{eq:gen_markov_cond}. One way to quantify the distance between the right hand side (r.h.s.) and the left hand side (l.h.s.) of Equation \eqref{eq:gen_markov_cond} is by using the Kullback-Leibler Divergence. In this case, testing Equation \eqref{eq:gen_markov_cond} becomes equivalent to test whether the expression
\begin{equation}\label{eq:transfer_entropy}
    T_{X \rightarrow Y}^{(k,l)}(\tau) = \sum_{Y_{t},\mathbf{Y}_{t-1}^{(k)},\mathbf{X}_{t-\tau}^{(l)}} p(Y_{t},\mathbf{Y}_{t-1}^{(k)},\mathbf{X}_{t-\tau}^{(l)})\log\frac{p(Y_{t}\vert \mathbf{Y}_{t-1}^{(k)},\mathbf{X}_{t-\tau}^{(l)})}{p(Y_{t}\vert \mathbf{Y}_{t-1}^{(k)})}
\end{equation}
is different from zero \citep{schreiber2000measuring}. Equation \eqref{eq:transfer_entropy} is known as \textit{transfer entropy} (TE). Note that $T_{X \rightarrow Y}^{(k,l)}(\tau=0)  \neq T_{Y \rightarrow X}^{(k,l)}(\tau=0)$, i.e., the transfer entropy is asymmetric as expected from a measure of predictive causality \citep{schreiber2000measuring}.

In principle to test causality one needs to include in the l.h.s. of Equation \eqref{eq:gen_markov_cond} all the information available in the universe at time $t-1$ and all the information available in the universe with the exception of $\mathbf{X}_{t-\tau}^{(l)}$ in the r.h.s \citep{pearl2009causality}. For controlled systems all the variables influencing the measurements of $Y$'s state are assumed to be known and the transfer entropy becomes measurable. This naturally set a limit to the application of such causal inference technique. For example, in the cases in which all the relevant variables are not known \textit{a priori}, we can still compute Equation \eqref{eq:transfer_entropy}, but it can be different from zero even though the interaction between $X$ and $Y$ is mediated by, e.g., a third variable $Z$ \citep[indirect causation; see e.g.][]{bossomaier2016transfer}. Thus, the predictive causality does not generally imply the true cause-effect relationship if the information of the whole set of relevant variables is not available. But it can be still used to measure lags between variables and directional coupling \citep{wibral2013measuring}.

From a numerical point of view, a key question is whether or not the values found for the transfer entropy are statistically significant. In our case, the critical value of the transfer entropy $T^{(*)}$ above which we can reject the null hypothesis is computed by generating surrogate time series satisfying Equation \eqref{eq:gen_markov_cond} and with the same statistical properties of $X$ and $Y$. In order to achieve this, we create surrogate trials by randomly shuffling the time series $X_{t-\tau}$. This allows us to estimate the distribution of the null hypothesis, to fix a confidence bound and to find the critical value $T^{(*)}$ which adapts to our dataset. The values such that $T_{X \rightarrow Y}^{(k,l)}(\tau) > T^{*}$ are considered statistically significant. 

\section{Results}
\label{results section}

\begin{figure}
    \centering
    \includegraphics[width=0.9\textwidth]{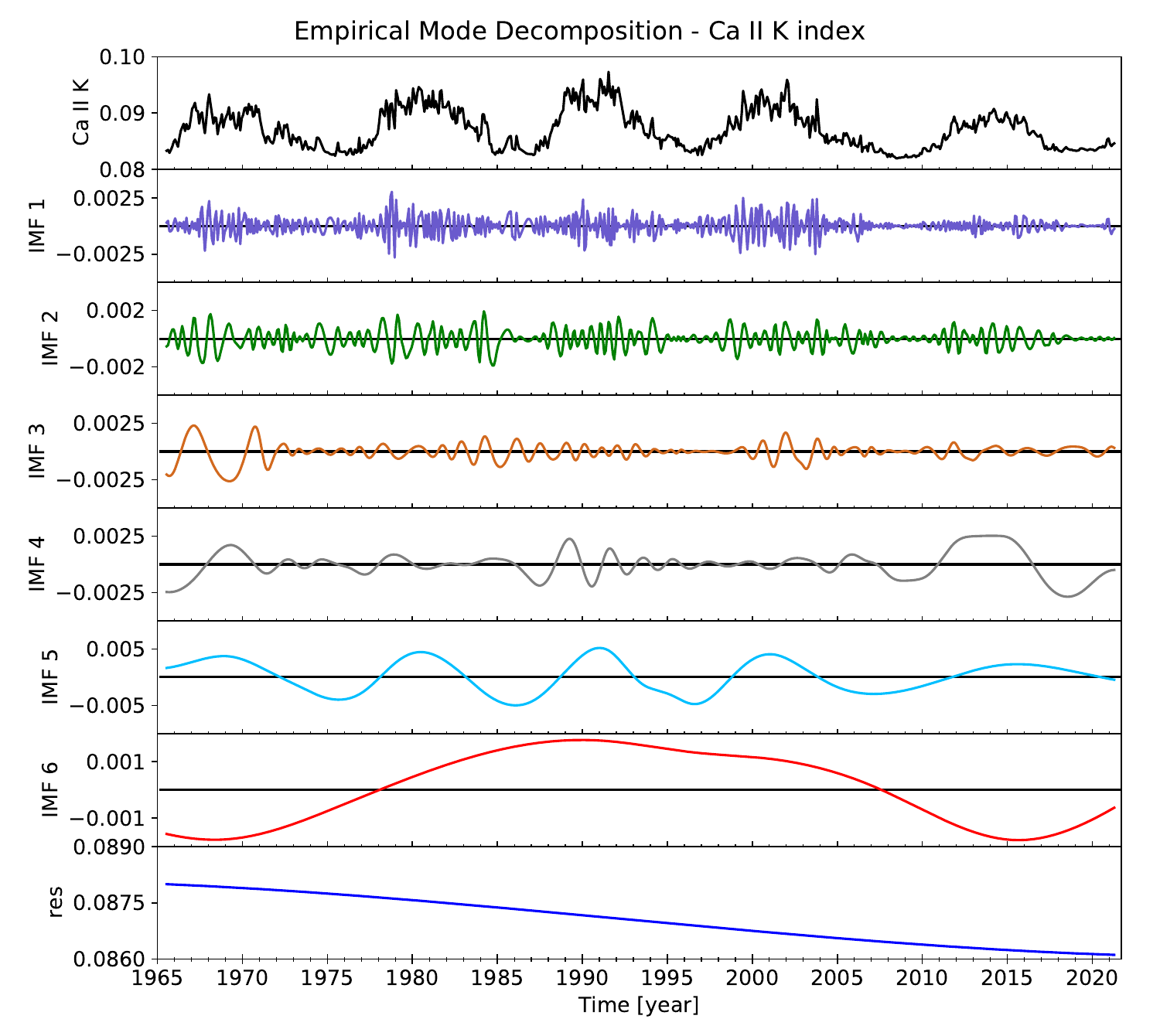}
    \caption{Empirical Mode Decomposition of Ca II K index. The top row shows the starting monthly means. The subsequent rows show the successive order IMFs, while the last row shows the residual signal.}
    \label{EMD_Ca II K}
\end{figure}

\begin{figure}
    \centering
    \includegraphics[width=0.9\textwidth]{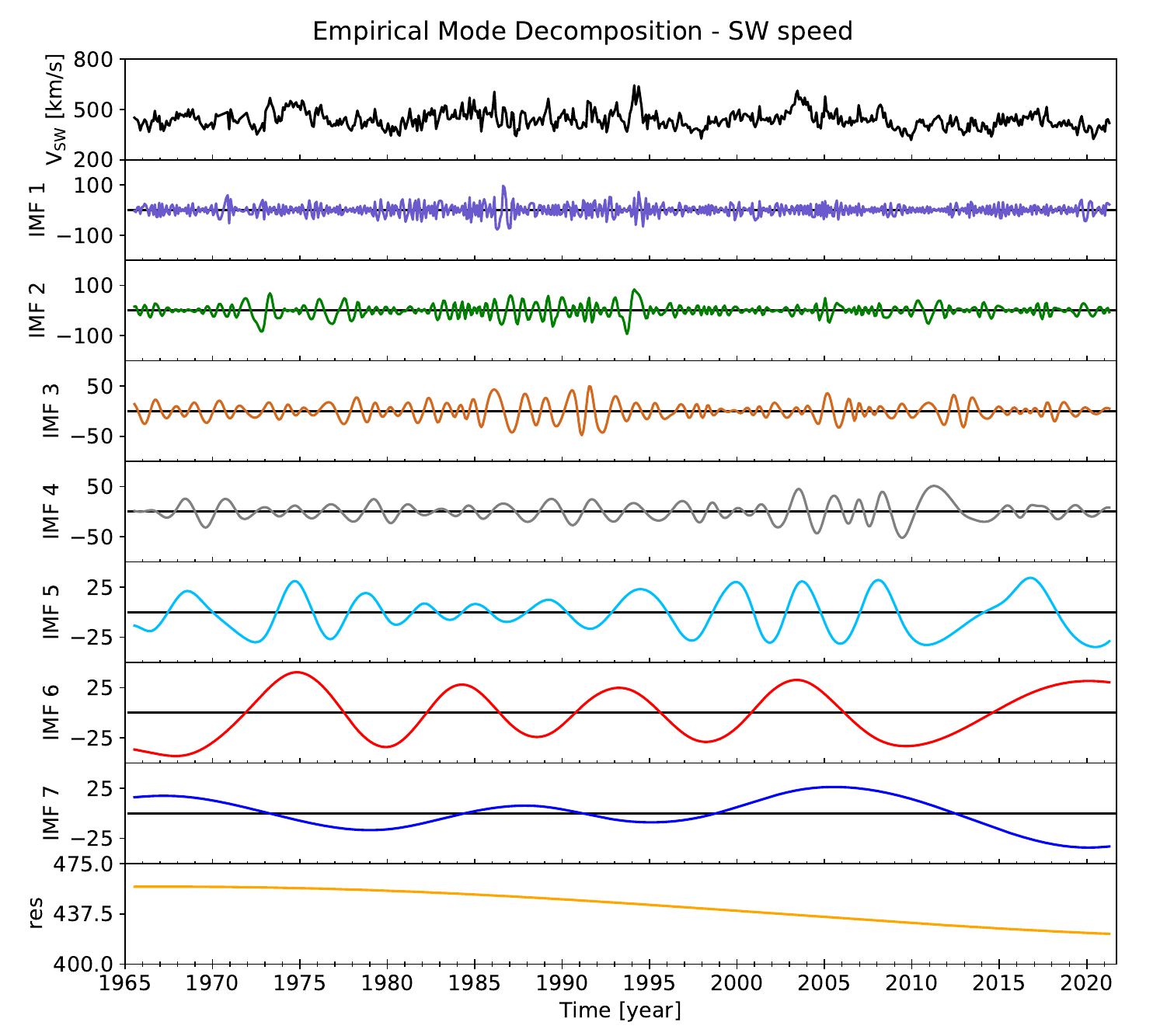}
    \caption{Empirical Mode Decomposition of solar wind speed. The top row shows the starting monthly means. The subsequent rows show the successive order IMFs, while the last row shows the residual signal.}
    \label{EMD solar wind speed}
\end{figure}

\begin{figure}
    \centering
    \includegraphics[width=0.9\textwidth]{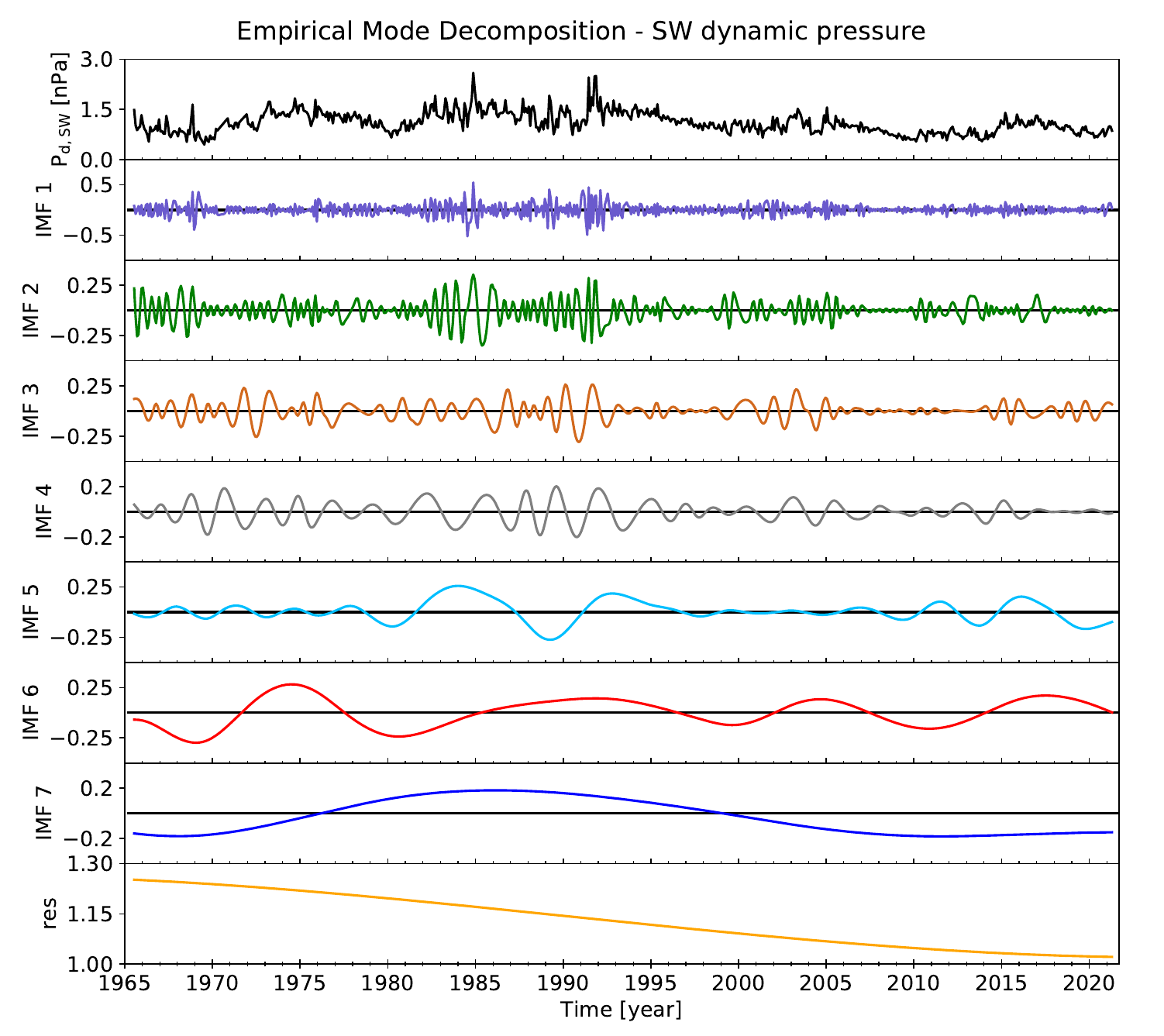}
    \caption{Empirical Mode Decomposition of solar wind dynamic pressure. The top row shows the starting monthly means. The subsequent rows show the successive order IMFs, while the last row shows the residual signal.}
    \label{EMD solar wind dynamic pressure}
\end{figure}

The results of Empirical Mode Decomposition are shown in Fig. \ref{EMD_Ca II K} for the Ca II K index, in Fig. \ref{EMD solar wind speed} for the solar wind speed and in Fig. \ref{EMD solar wind dynamic pressure} for the solar wind dynamic pressure. The mode decomposition generates 6 IMFs for the Ca II K index, 7 IMFs for the solar wind speed and 7 IMFs for the solar wind dynamic pressure.

The characteristic time scales (or the average period) of the IMFs obtained with the EMD are shown, for each signal, in Table \ref{table IMFs}.
\begin{table}[]
    \caption{Characteristic time scales of the extracted IMFs for each signal.}
    \label{table IMFs}
    \centering
    \begin{tabular}{c|c|c|c}
    \hline
    \hline
        IMF \# & \multicolumn{3}{c}{Mean period [years]} \\
       \hline
        & Ca II K index & SW speed & SW dynamic pressure \\
       \hline
        1 & 0.36 & 0.31 & 0.32 \\
        2 & 0.65 & 0.59 & 0.54 \\
        3 & 1.36 & 1.06 & 1.18 \\
        4 & 3.58 & 2.13 & 2.26 \\
        5 & 12.38 & 4.84 & 5.17 \\
        6 & 51.43 & 11.17 & 13.88 \\
        7 & - & 19.84 & 51.45 \\
        \hline
        \hline
    \end{tabular}
\end{table}
The Ca II K index, as expected, displays an intrinsic component (IMF 5) related to the 11-year solar activity cycle, in particular with a mean period of 12.4 years. Also the solar wind speed and dynamic pressure display a component at solar cycle time scales. For both parameters it is the IMF 6, corresponding to a mean period of 11.2 years for the speed and 13.9 years for the dynamic pressure.
It is interesting to notice that for solar wind speed and dynamic pressure we obtain a mode with a quasi-biennial periodicity (IMF 4 in both cases), while this is not true for Ca II K index.

\begin{figure}
    \centering
    \includegraphics[width=\textwidth]{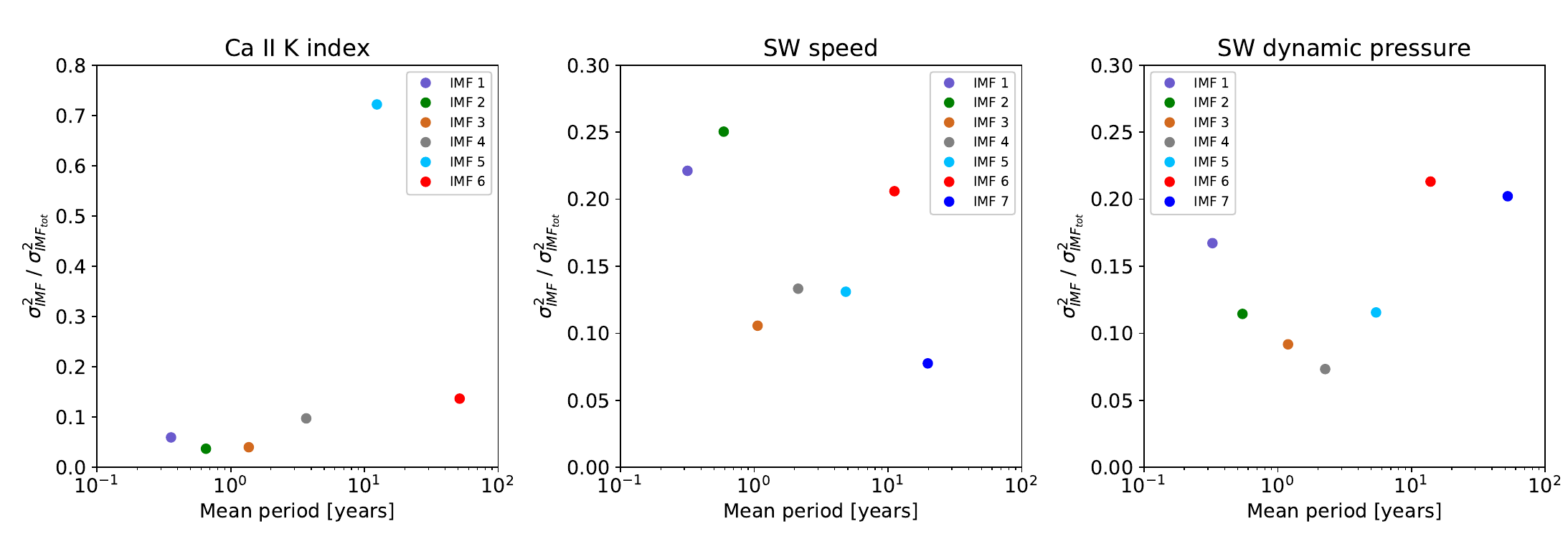}
    \caption{IMF variance weighted on the IMFs total variance as a function of the IMF mean period. The subplots are for Ca II K index (left), solar wind speed (center) and solar wind dynamic pressure (right).}
    \label{Varianze IMFs pesate}
\end{figure}

In order to understand which IMF has the highest contribute to the overall variability of the signal, it is possible to compute for each IMF a weighted variance as it follows. The value of the variance of each IMF ($\mathrm{\sigma^{2}_{IMF}}$) is normalized to that of the total signal ($\mathrm{\sigma^{2}_{IMF_{tot}}}$), obtained by summing the contribution of all the IMFs but excluding the residual term, and plotted as a function of the mean period of the IMF itself. Such values are shown for the Ca II K index, the solar wind speed and the solar wind dynamic pressure in Fig. \ref{Varianze IMFs pesate}.
We can use this information to quantify the contribution of each IMF to the overall observed behaviour. We focus for our analysis on the IMFs with a mean period over one year. In the case of Ca II K index (left panel of Fig. \ref{Varianze IMFs pesate}), the greatest weighted variance is from IMF 5, the one related to the 11-year cycle. 
In the case of the solar dynamic pressure (right panel of Fig. \ref{Varianze IMFs pesate}), the major contribution is from IMF 6, once again the one corresponding to the solar cycle time scales. The same is true for the solar wind speed (central panel of Fig. \ref{Varianze IMFs pesate}), for which over the yearly time scale the highest contribution is from IMF 6.

\begin{figure}
    \centering
    \includegraphics[width=\textwidth]{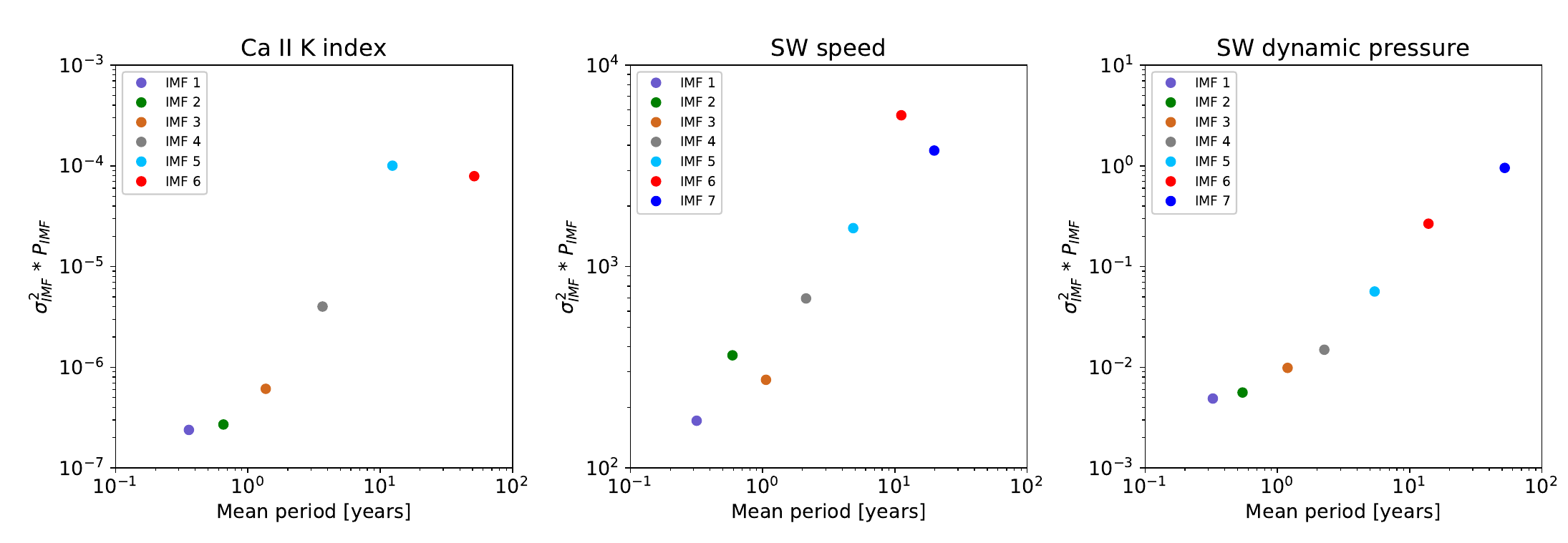}
    \caption{IMF variance multiplied by the mean period as a function of the IMF mean period (log-log scale). The subplots are for Ca II K index (left), solar wind speed (center) and solar wind dynamic pressure (right).}
    \label{Varianze IMFs per periodo}
\end{figure}

To further investigate the results of the decomposition, it is possible to look at how the power is distributed among the IMFs, searching for the presence of possible power laws. This can be done by plotting, for each IMF, the value of the variance ($\mathrm{\sigma^{2}_{IMF}}$) multiplied by the corresponding mean period ($\mathrm{P_{IMF}}$), as a function of the mean period of the IMF itself (log-log scale). This is the analogous of a spectral density. The results are shown for Ca II K index, solar wind speed and dynamic pressure in Fig. \ref{Varianze IMFs per periodo}. It is possible to notice the presence of a power law, characterized by increasing intensity from yearly to solar cycle scales, thus extending over at least one decade, in all the signals. Although in this time range the power law is clearly evident, the behaviour at very high and very low frequencies remains uncertain due to the limited data points available in the plots.

\begin{figure}
    \centering
    \includegraphics[width=0.8\textwidth]{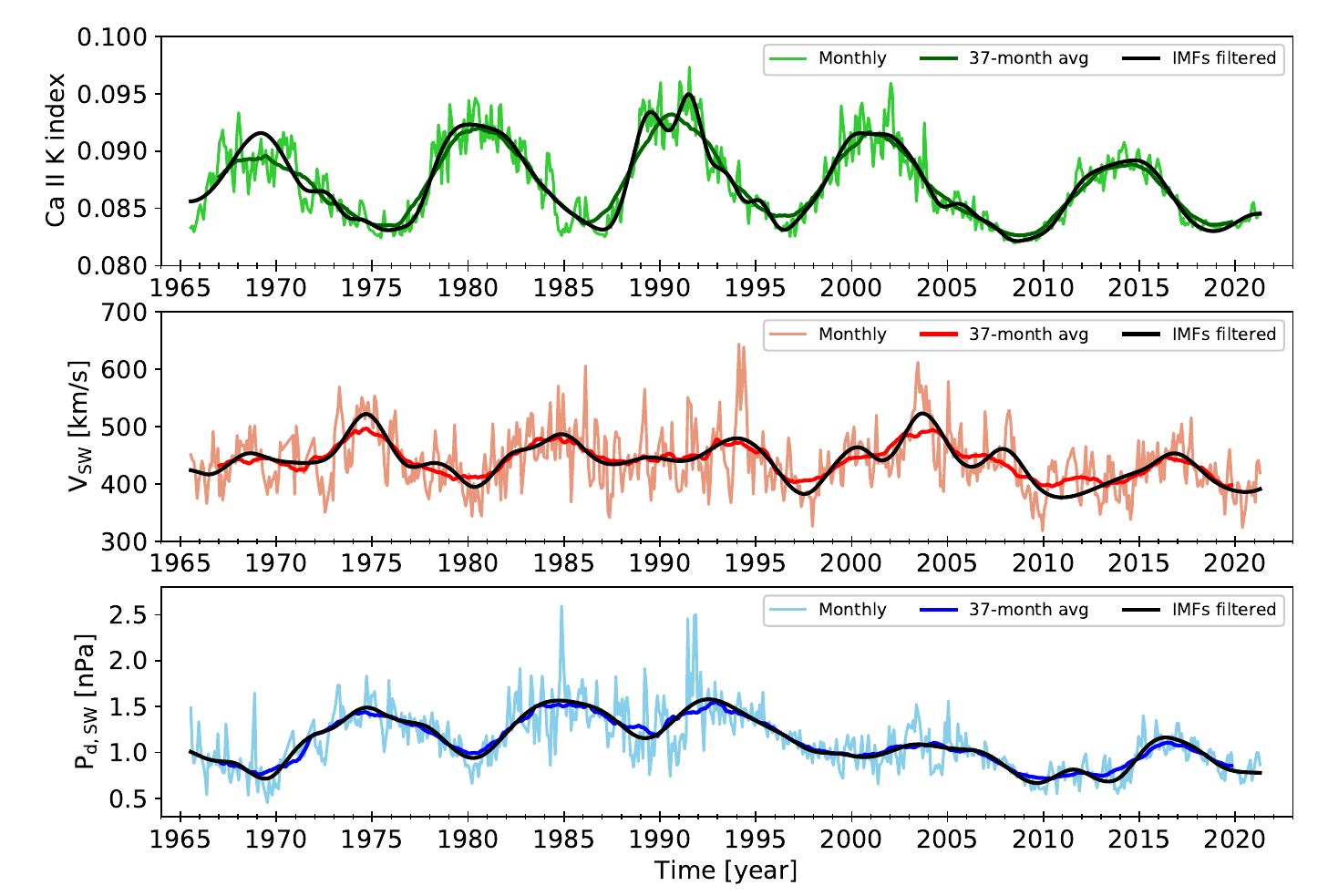}
    \caption{Monthly averages with superimposed 37-month moving averages and IMFs filtered signals for Ca II K index (top), solar wind speed (middle) and solar wind dynamic pressure (bottom).}
    \label{Confronto medie segnali}
\end{figure}

Once the signals have been decomposed via EMD as shown above, it becomes possible to filter the time series of the three parameters by means of the obtained IMFs. To this scope, we subtracted from the monthly time series of Ca II K index, solar wind speed and dynamic pressure the contribution of the IMFs with mean periods smaller than 3 years. This criterion is chosen in order to be consistent and to compare the results with the 37-month filtering previously applied in \cite{Reda2023}. In particular, for the Ca II K index the contribution from IMFs 1-3 have been subtracted, while for both solar wind parameters we subtract the IMFs 1-4. The comparison between monthly means, 37-month averages and IMFs filtered data for the three signals is shown in Fig. \ref{Confronto medie segnali}. As it can be seen, the signals obtained by filtering the high-frequency IMFs are consistent with the 37-month moving averages, but they seem to better follow the behaviour of the monthly means with respect to the latter. Indeed, the signals filtered by means of the IMFs retain more information, such as the double solar cycle peak visible in Ca II K index (top panel of Fig. \ref{Confronto medie segnali}).

\begin{figure}
    \centering
    \includegraphics[width=0.8\textwidth]{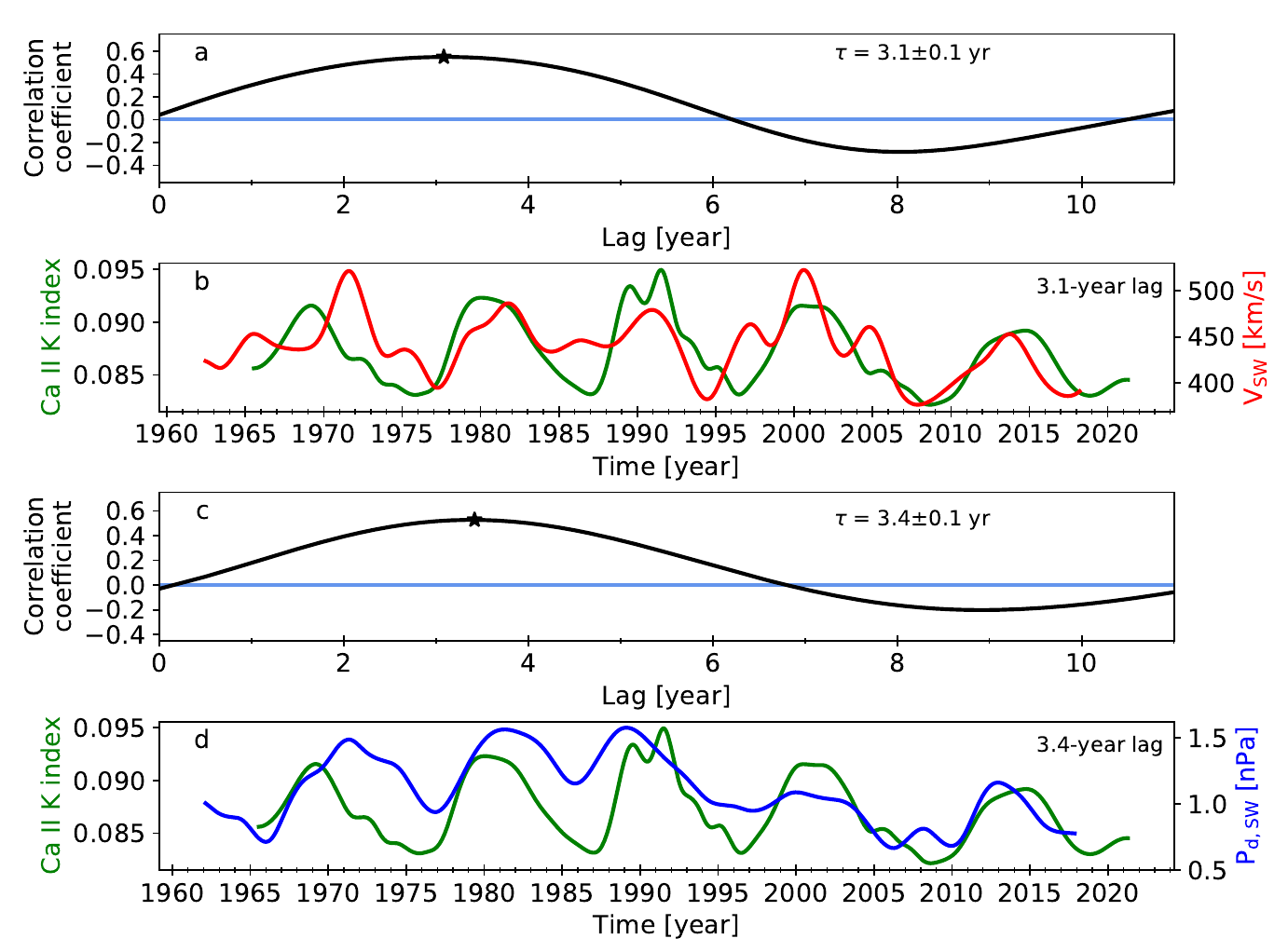}
    \caption{Cross correlation of the IMFs filtered signals. a) Cross-correlation between Ca II K index and solar wind speed; b) Comparison of Ca II K index (green) with solar wind speed (red) shifted backward by 3.1 years; c) Cross-correlation between Ca II K index and solar wind dynamic pressure; d) Comparison of Ca II K index with solar wind dynamic pressure (blue) shifted backward by 3.4 years.}
    \label{Cross correlation}
\end{figure}

\begin{figure}
    \centering
    \includegraphics[width=0.49\textwidth]{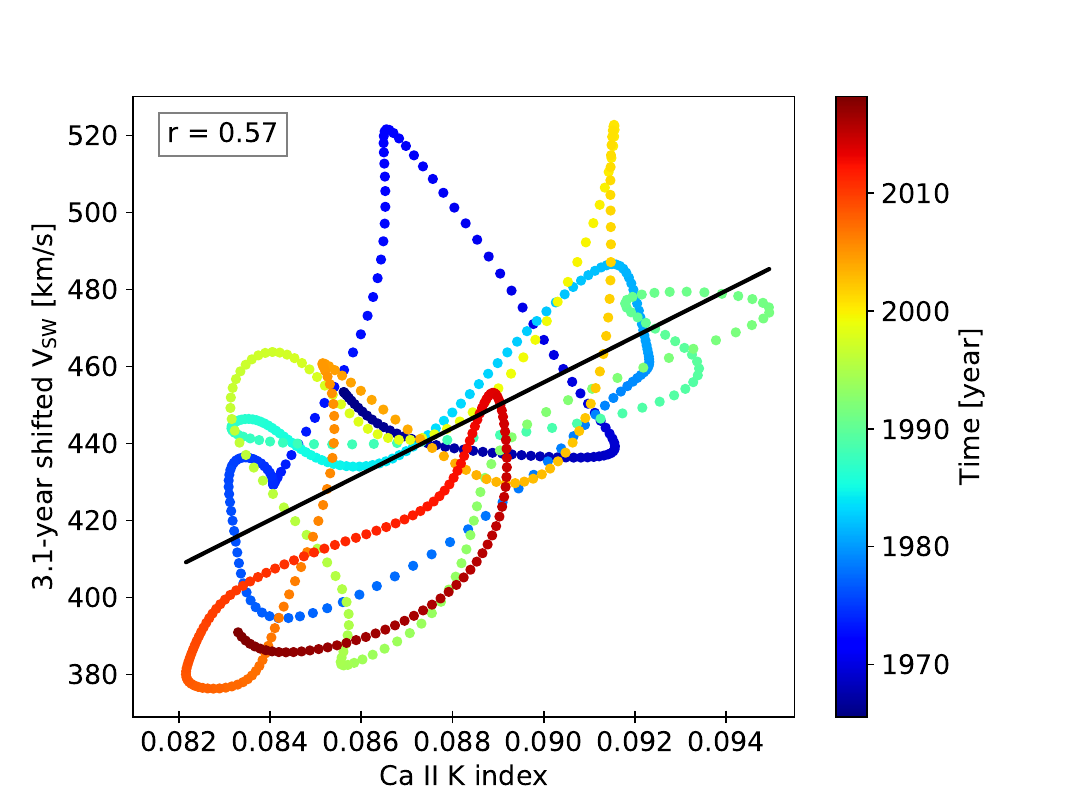}
    \includegraphics[width=0.49\textwidth]{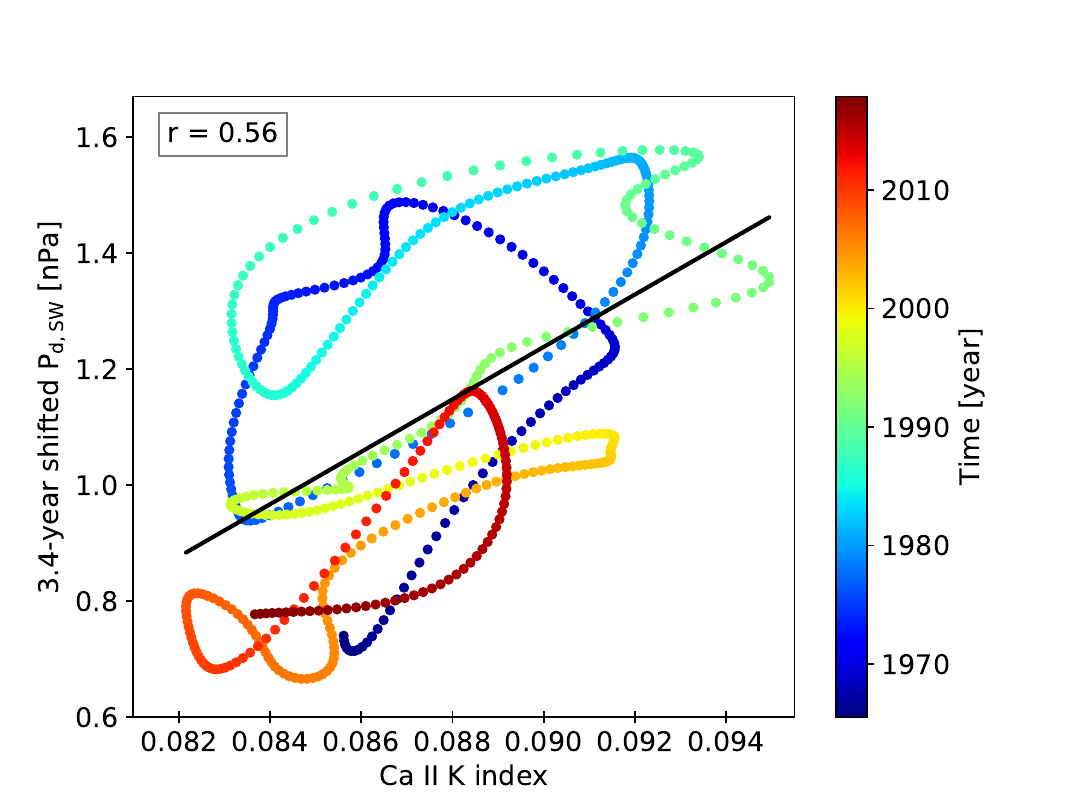}
    \caption{Scatter plot showing the relationship of Ca II K index with solar wind speed (left panel) and solar wind dynamic pressure (right panel) once shifted by the time lags found with the cross-correlation analysis. In both panels the color map shows how the relation changes with time, while the black line shows the best linear fit to the data points. The Pearson correlation coefficients are reported on the upper-left.}
    \label{Scatter color maps}
\end{figure}

According to the analysis performed in \cite{Reda2023}, once the high-frequency components of the signals have been filtered out, it is possible to investigate the time lag between them. To this scope, we use here a cross-correlation analysis considering only positive delay of the solar wind parameters with respect to the activity of the Sun (Ca II K index here). The results of the cross-correlation analysis are shown in Fig. \ref{Cross correlation}. The maximum correlation between Ca II K index and solar wind speed occurs at a time lag of $3.1 \pm 0.1$ yr, in agreement with the result of $3.2 \pm 0.1$ yr found in \cite{Reda2023} by using 37-month averaged data. The maximum correlation of Ca II K index with solar wind dynamic pressure, instead, is found for a time lag of $3.4 \pm 0.1$ yr. This value is in agreement, within the confidence intervals, with the values found in \citet{Reda2023} with cross-correlation ($3.6 \pm 0.1$ yr) and mutual information ($3.4 \pm 0.1$ yr). These findings strengthen the analysis, as the results do not depend on the technique adopted to filter out the high-frequency components.

The scatter plots of Fig. \ref{Scatter color maps} show the relation of Ca II K index with solar wind speed and solar wind dynamic pressure respectively, once the time lags from the cross-correlation analysis are taken into account. In both figures the black lines show the best linear fits to the data points. The Pearson’s correlation coefficient is r = 0.57 in the case of the speed and r = 0.56 in the case of the dynamic pressure, indicating in both cases a positive moderate correlation. For the case of Ca II K index with solar wind dynamic pressure the correlation coefficient is almost equal to that found in \cite{Reda2023} by using 37-month averaged data (r = 0.57), while in the case of solar wind speed it is smaller compared to the value they found (r = 0.65). 

\begin{figure}
    \centering
    \includegraphics[width=0.49\textwidth]{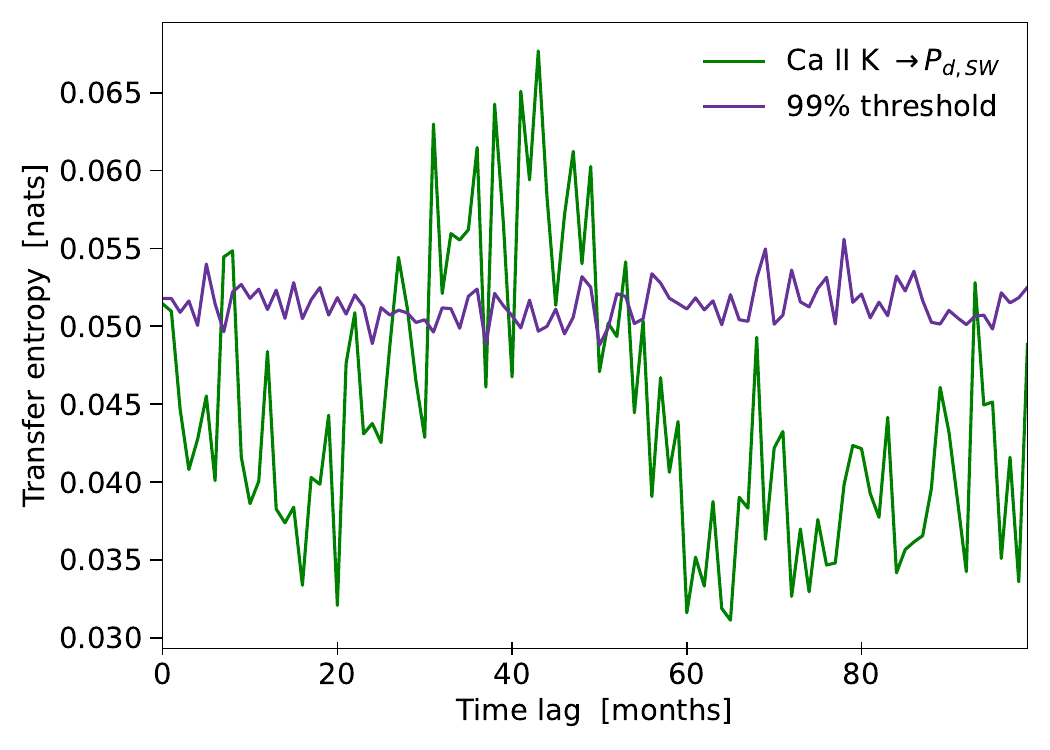}
    \includegraphics[width=0.49\textwidth]{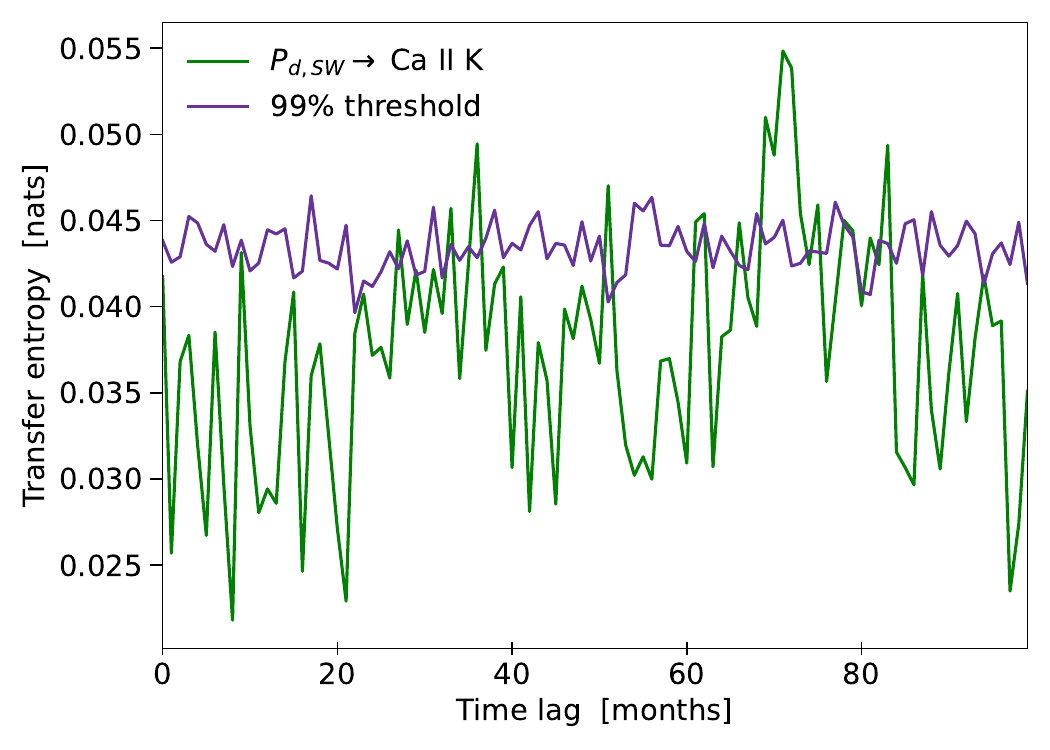}
    \includegraphics[width=0.49\textwidth]{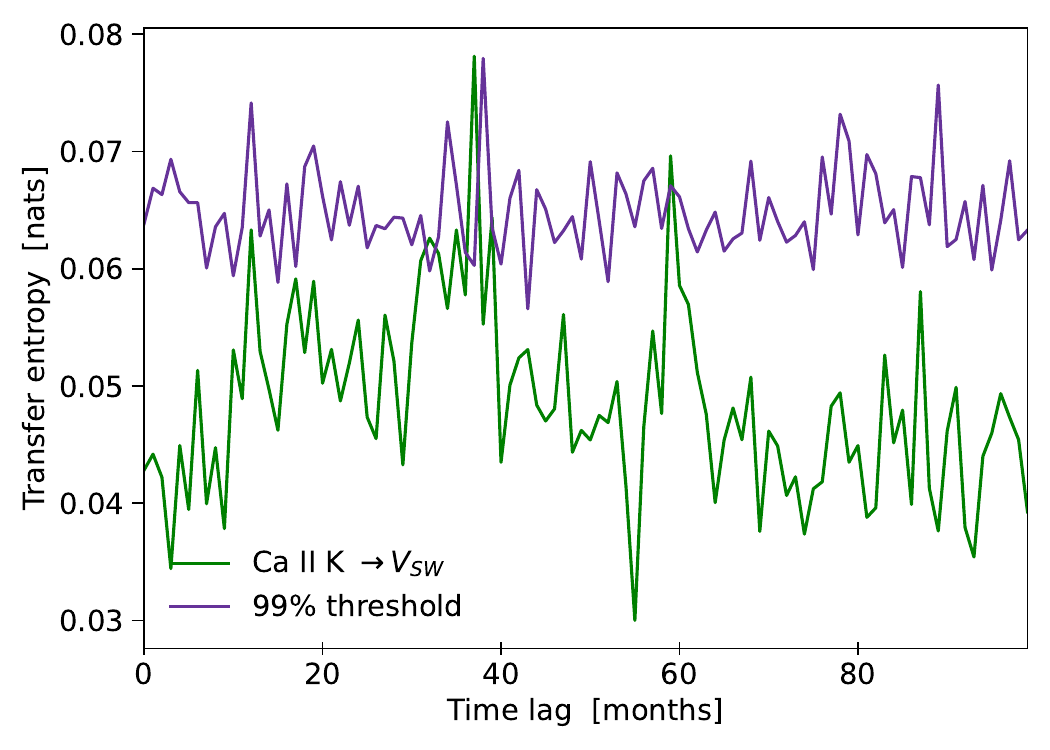}
    \includegraphics[width=0.49\textwidth]{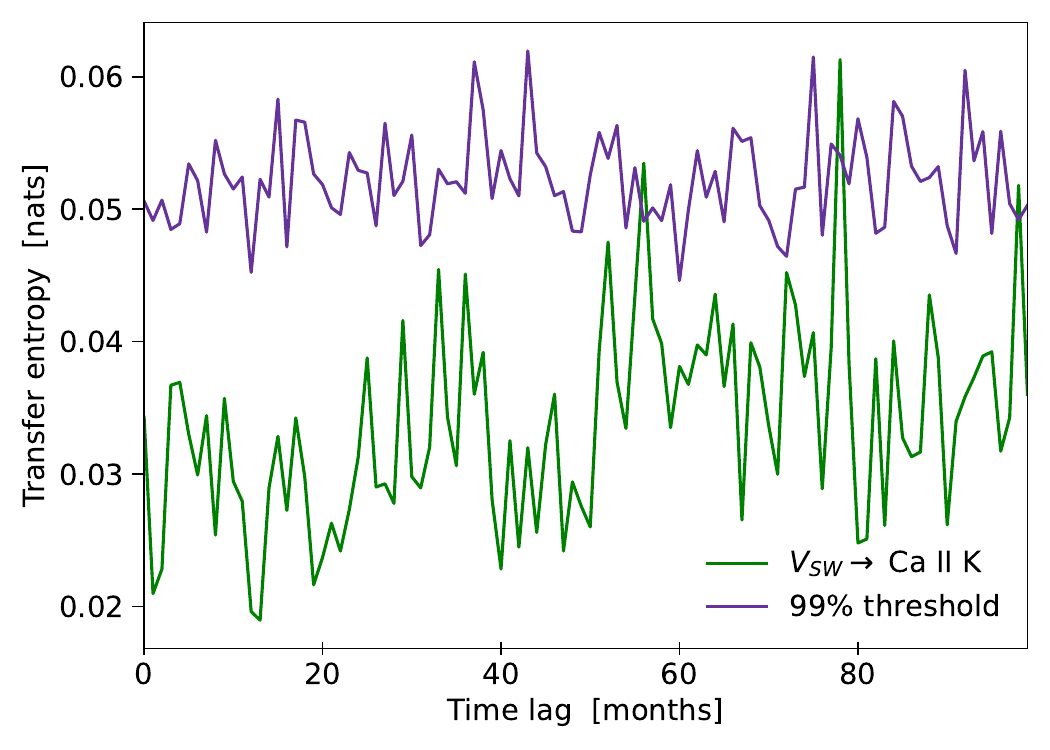}
    \caption{Transfer entropy from Ca II K index to solar wind dynamic pressure and speed (top and bottom left panels, respectively) and \textit{vice-versa} (top and bottom right panels, respectively), computed by using monthly averaged data. The purple line marks the 99\% significance threshold obtained from the analysis of surrogate time series data.}
    \label{fig:te_pressure}
\end{figure}

In order to assess in a stronger framework the results obtained with the present analysis, we compute the transfer entropy by directly employing the monthly averaged data, without applying any filter. This approach allows us to investigate higher-order correlation between data, i.e., predictive causality as explained in the previous section. In Fig. \ref{fig:te_pressure} we show the information flow, as measured by the transfer entropy, from Ca II K index to $\mathrm{P_{d,SW}}$ (top-left panel) and \textit{vice-versa} (top-right panel). In both cases the purple line shows the empirical threshold of 99\% arising from the analysis of 500 surrogate time series data (see Section \ref{transfer entropy section}). The information flow from the Ca II K index to $\mathrm{P_{d,SW}}$ exhibits a statistically significant structure/enhancement between $\sim$25 and $\sim$ 50 months. The maximum of the transfer entropy is at 43 months ($\simeq$ 3.6-year), while the mean of the interval (assuming a symmetric peak) is at 37.5 months ($\simeq$ 3.1-year). 
Both lags are comparable with the results found with the cross correlation analysis in this work, but also in agreement with the findings by \cite{Reda2023}. The latter result highlights that the correlation found is not simply due to the synchronization of the time series peaks, but it means there is a predictive link between Ca II K index and the solar wind dynamic pressure, suggesting a certain degree of causation. 

On the other hand, the information flow in the reversed case (top-right panel of Fig. \ref{fig:te_pressure}) is always below the 99\% threshold with the exception of a peak at 71 months ($\simeq$ 5.9-year; comparable with the distance between solar maximum and solar minimum). We interpret this finding as due to redundancies/periodicities induced by the solar cycle. In order to test quantitatively this hypothesis, in principle Eq. \eqref{eq:transfer_entropy} should be computed by using $k>1$. However, this is not reliable with only 670 data points.

The bottom panels of Figure \ref{fig:te_pressure} show the results of the transfer entropy analysis for the solar wind speed. In both directions, i.e., from Ca II K index to $\mathrm{V_{SW}}$ (bottom-left panel) and \textit{vice-versa} (bottom-right panel), there are no strong evidences about the information flow. The exceedances of the threshold at lags of 37 months ($\simeq$ 3.1-year) and 60 months ($\simeq$ 5.0-year) from  Ca II K to $\mathrm{V_{SW}}$, and at 78 months ($\simeq$ 6.5-year) from $\mathrm{V_{SW}}$ to Ca II K, are here interpreted as a fluctuation due to sampling effects. However the peak at 3.1-year is consistent with the structures found in the analysis of solar wind dynamic pressure and with the results recently found in \cite{Reda2023}.  

Note that in general the results of the transfer entropy analysis are noisy. This is due to the fact that we have only 670 data points, so that the estimation of high-dimensional transition probabilities is prone to fluctuations. To mitigate this effect, the estimation of transition probabilities must be performed by reducing the number of bins. In our case, the best trade-off between correct sampling and resolution (i.e., in order to have filled bins) is to choose less than 10 bins per dimension. 
Aware of this technical problem, we interpret threshold exceedances of single-point structures as fluctuations that are not statistically significant.

\section{Discussion and conclusions}
\label{discussion and conclusions section}
Starting from the monthly averages of a physical proxy of the solar activity (the Ca II K index) and solar wind parameters, we investigate in this work their relationship on Space Climate scales. To this scope, we take advantage of the Hilbert-Huang Transform. This method allows to decompose the starting signals into several modes and to obtain for each of them the instantaneous frequency and hence the mean characteristic time scale. 
Looking at how the energy is distributed among the time scales, we find a quite similar behaviour for all the signals between annual and solar cycle scales, characterized by an increasing power. Concerning the behaviour at lower and higher time scales, instead, it is not possible to draw conclusions here.

The advantage of the HHT is that the EMD makes possible to filter out the noisy high-frequency components, which are not of interest for the purpose of this work. The time lags we find between Ca II K index and both solar wind speed (3.1-year) and dynamic pressure (3.4-year), after subtracting the contribution at scales smaller than 3 years, are consistent with the results previously obtained by \cite{Reda2023} using a 37-month smoothing on the same dataset.

However, the presence of a correlation with a time delay does not ensure a cause-effect relation, as well as the presence of mutual information peak does not guarantee the nature of the directional coupling. For this reason, in order to further investigate these results, we use the transfer entropy as predictive causality test for higher-order correlation between data. The results from transfer entropy analysis suggest the presence of statistically significant structures from Ca II K index to solar wind dynamic pressure, with a peak at time lag of 3.6-year, once again in agreement with the time lag found by \cite{Reda2023}.
This finding confirms that the knowledge of past values of the Ca II K index gives information about the future state of the solar wind dynamic pressure. Indeed, the cross correlation analysis is based on co-variation of data and it is naturally stronger when peaks are synchronized, while the transfer entropy is based on (temporal) transition probabilities between the states and it is a dynamical and time-asymmetric concept. As reported by \cite{wibral2013measuring}, the time-delays found from cross correlation analysis and transfer entropy analysis are not necessarily the same.

Considering the information flow from Ca II K index to solar wind speed, the transfer entropy shows a single peak that exceeds the 99\% threshold, at time lag of 3.1-year. As in the case of the dynamic pressure, this results is consistent with the result recently found by \cite{Reda2023}. However, because it is a single peak it may also be interpreted as a fluctuation. 

Our results suggest that over the last five solar cycles there is a better information flow from Ca II K index to solar wind dynamic pressure than from Ca II K index to solar wind speed. Since the dynamic pressure depends both on the speed and the density of the solar wind, thus making it an energy related parameter, we interpret this result as a phenomenon connected to energy transfer processes from the Sun to the heliosphere. The former result could be of interest to build up a predictive model in a space climate context.

We remark that, although the results found with the transfer entropy analysis are in agreement with previous findings, further work is needed to assess the causal relationship. Indeed, due to technical limitations (e.g., few data points) we are not able to investigate thoroughly the statistical significance of our results. However, we emphasize that, at the moment, the application of the transfer entropy is promising and may be extremely helpful in the future to disentangle the (non-linear) causal relations between the solar activity and the solar wind at space climate scales. 
Dataset with a higher time resolution, thus with more data points, will be considered for a future analysis in order to verify this hypothesis and to confirm the result obtained via the transfer entropy presented in this work.

\backmatter

\bmhead{Acknowledgments}

The results presented in this article are based on Ca II K index composite data, which are freely accessible at the SOLIS website (\url{https://solis.nso.edu/0/iss/}). SOLIS is managed by the National Solar Observatory. The Mg II composite from the University of Bremen is available at \url{https://www.iup.uni-bremen.de/UVSAT/Datasets/mgii}. The OMNI data have been downloaded from the Space Physics Data Facility (SPDF) Coordinate Data Analysis Web (CDAWeb) at \url{https://cdaweb.gsfc.nasa.gov/}. The authors acknowledge all the persons who made the availability of the mentioned data possible. 
R.R. acknowledges the support from the European Union’s Horizon 2020 research and innovation program under grant agreement No. 824135 (SOLARNET).
L.G. acknowledges the support from PON - FESR REACT-EU MUR DM 1062.
T.A. acknowledges the "Bando per il finanziamento di progetti di Ricerca Fondamentale 2022" of the Italian National Institute for Astrophysics (INAF) - Mini Grant: "The predictable chaos of Space Weather events".

\section*{Declarations}

The authors declare no conflict of interest.


\bibliography{sn-bibliography}


\end{document}